\documentclass[prd,showpacs,twocolumn]{revtex4}
\ifx\undefined\fiverm
     \font\fiverm=cmr5
\fi

\input{prepictex}
\input{pictex}
\input{postpictex}

\begin{document}
\author{Tarun Biswas}
\title{Physical Interpretation of Coordinates for the Schwarzschild Metric}
\email{biswast@newpaltz.edu}
\affiliation{State University of New York at New Paltz, \\ New Paltz,  NY 12561, USA.}
\date{\today}
\begin{abstract}
Since its first introduction, the Schwarzschild metric has been written in various coordinate
systems. This has been done primarily to understand the nature of the coordinate singularity
at the event horizon. However, very often, the mathematics of a coordinate system does
not provide a clear physical interpretation. In Schwarzschild's original work, the origin of the
radial coordinate was at the event horizon. Hence, there was no black hole. The generally accepted
current definition of the radial coordinate has an origin beyond the horizon. This necessitates
the discussion of black holes. Here, some well-known and some not-so-well-known coordinate
systems will be visited in search of a physical interpretation. It will be noted that they all agree
at large radial distances. However, the location of the origin of the radial coordinate can be
different for different systems. Some ``natural'' coordinate systems can exclude the entirety of the
interior of a black hole from the physical manifold. Such coordinate systems maybe physically
more acceptable as they avoid the issue of the metric signature change across the horizon.
Mathematically, the metric signature alone can distinguish time from space coordinates. Hence,
coordinate systems that include the interior of the black hole need to switch the physical
meanings of time and one of the space coordinates.
\end{abstract}
\maketitle
\section{Introduction}
It is customary to use the standard spherical polar coordinates $(t,r,\theta,\phi)$ to describe the
spherically symmetric vacuum solution to Einstein's equations. This coordinate system is meant for
a ${\cal R}^{4}$ topology. However, the solution to Einstein's equations distorts not only the
flatness of ${\cal R}^{4}$ but also its topology. As a result, the spatial ${\cal R}^{3}$ part
needs to have a region around the origin amputated to provide the physical manifold. Standard
practice is to exclude only the singular point at the origin ($r=0$). However, here it will be
argued that a finite spherical region needs to be excluded to maintain the physical nature of the
manifold. In either case, what is
left of the spatial part has the topology of a semi-infinite cylinder.

The Schwarzschild line element in standard polar coordinates is given as
\begin{equation}
	d\tau^{2}=\left(1-\frac{r_{s}}{r}\right)dt^{2}-\left(1-\frac{r_{s}}{r}\right)^{-1}dr^{2}
	-r^{2}d\Omega^{2},
\end{equation}
where
\begin{equation}
	d\Omega^{2}=d\theta^{2}+\sin^{2}\theta d\phi^{2},
\end{equation}
the speed of light $c=1$, $r_{s}=2GM/c^{2}$, $G$ is the universal gravitational constant
and $M$ is the mass of the source. An observer starting at a point $r=r_{0}>r_{s}$ and 
falling in a radial direction in this metric is known to take infinite coordinate 
time $t$ to reach $r=r_{s}$ (the Schwarzschild radius)\cite{misner}.
Hence, once the observer has fallen past $r_{s}$, the coordinate time $t$ exceeds $\infty$
which is meaningless. This might render the above line element itself meaningless for such
an observer as $t$ is used in its definition. Although the observer takes a finite amount
of time $\tau$ to reach $r=r_{s}$ in his/her own frame, it is not clear what he/she ``sees''
for $r<r_{s}$ as the above metric can no longer be used. There is no real coordinate transformation
available between $\tau$ and $t$ for $r<r_{s}$. This is why, if the above metric is forced for
$r<r_{s}$, the metric signature changes. In other words, $t$ needs to be imaginary!

In this light, it is useful to consider coordinate systems that exclude the $r\leq r_{s}$
region (interior of a black hole) from the physical manifold in a ``natural'' manner. First
I shall review some well-known coordinate systems that include the interior of the black
hole (Eddington-Finkelstein and Kruskal). Then I shall discuss some others that effectively
exclude the interior of the black hole. Among these will be the somewhat well-known isotropic
coordinates. But there are several others that produce the same effect in a much simpler fashion.
The original system used by Schwarzschild is among them\cite{schwarz, schwarzt}. An indirect
reason for excluding the interior of a black hole has been seen recently in some quantum
effects\cite{vachas}.

\section{The Eddington-Finkelstein coordinates}
First, let us look at the ingoing Eddington-Finkelstein coordinates\cite{misner}. Here the
coordinate $t$ is replaced by the null coordinate $V$ given by
\begin{equation}
	V=t+r+r_{s}\ln|r/r_{s}-1|.
\end{equation}
Thus, the line element is given by
\begin{equation}
	d\tau^{2}=(1-r_{s}/r)dV^{2}-2dVdr-r^{2}d\Omega^{2}.
\end{equation}
This removes the singularity in the metric component at $r=r_{s}$ by using a 
coordinate transformation
that is singular at the same point. Hence, it is often argued that the $(V,r)$ coordinates represent
reality better at the Schwarzschild horizon ($r=r_{s}$) than the original $(t,r)$ coordinates.
However, $t$ is directly measured by some observer's clock while $V$, being a null coordinate,
has no direct observational meaning from any observer's point of view. 
Besides, removing the singularity in a metric component through
a coordinate transformation is only a cosmetic advantage as the components of the curvature
tensor are well-behaved in either coordinate system. So, on physical grounds, $t$ might still
be the ``better'' coordinate.

The real problem of the Schwarzschild metric is not the coordinate singularity, but the
metric signature in the interior of the black hole. It is the metric signature alone that
mathematically distinguishes space from time. Physically, space and time are very different.
But the mathematics of relativity makes them look very similar. The only mathematical tag
that tells us which is which is the metric signature. Within the black hole this metric signature
is switched for $t$ and $r$ and this plays havoc with physical interpretation. Hence, it
would be useful to eliminate the interior of a black hole in a ``natural'' way if possible.

The ingoing Eddington-Finkelstein coordinates are often used to show that objects can fall
into a black hole but they cannot ``fall'' out. This is a bit perplexing as the original
metric is time reversal symmetric! To reconstruct this original symmetry, one has to look
at the other Eddington-Finkelstein coordinate system -- the outgoing one. For this, one defines
\begin{equation}
	U=t-r-r_{s}\ln|r/r_{s}-1|.
\end{equation}
The resulting form for the line element is
\begin{equation}
	d\tau^{2}=(1-r_{s}/r)dU^{2}+2dUdr-r^{2}d\Omega^{2}.
\end{equation}
This describes the case of objects that ``fall'' out of a black hole.

\section{The Kruskal coordinates}
The Kruskal coordinates\cite{misner} $u$ and $v$ are defined to replace $r$ and $t$ of the
standard coordinates. Like the Eddington-Finkelstein coordinates, they effectively remove
the singularity of the metric components by using a singular coordinate transformation.
But the Kruskal coordinates have no null coordinates. This might make physical interpretation
somewhat easier. However, the switching of the metric signature within the black hole is
still a problem. Also, the Kruskal coordinates cover the physical space twice! This is an
added challenge to their physical interpretation. The coordinates $u$ and $v$ are defined 
in four patches:
\begin{eqnarray}
\mbox{I}\left\{
\begin{array}{ll}
u=(r/r_{s}-1)^{1/2}e^{r/2r_{s}}\cosh(t/2r_{s}), & \mbox{for } r>r_{s} \\
v=(r/r_{s}-1)^{1/2}e^{r/2r_{s}}\sinh(t/2r_{s}), & \mbox{for } r>r_{s}
\end{array}
\right. & & \\
\mbox{II}\left\{
\begin{array}{ll}
u=(1-r/r_{s})^{1/2}e^{r/2r_{s}}\sinh(t/2r_{s}), & \mbox{for } r<r_{s} \\
v=(1-r/r_{s})^{1/2}e^{r/2r_{s}}\cosh(t/2r_{s}), & \mbox{for } r<r_{s}
\end{array}
\right. & & \\
\mbox{III}\left\{
\begin{array}{ll}
u=-(r/r_{s}-1)^{1/2}e^{r/2r_{s}}\cosh(t/2r_{s}), & \mbox{for } r>r_{s} \\
v=-(r/r_{s}-1)^{1/2}e^{r/2r_{s}}\sinh(t/2r_{s}), & \mbox{for } r>r_{s}
\end{array}
\right. & & \\
\mbox{IV}\left\{
\begin{array}{ll}
u=-(1-r/r_{s})^{1/2}e^{r/2r_{s}}\sinh(t/2r_{s}), & \mbox{for } r<r_{s} \\
v=-(1-r/r_{s})^{1/2}e^{r/2r_{s}}\cosh(t/2r_{s}), & \mbox{for } r<r_{s}
\end{array}
\right. & &
\end{eqnarray}
The resulting form for the line element is
\begin{equation}
	d\tau^{2}=(4r_{s}^{3}/r)e^{-r/r_{s}}(dv^{2}-du^{2})-r^{2}d\Omega^{2},
\end{equation}
where $r$ is implicitly a function of $u$ and $v$.

The metric signature problem (discussed in the last section) is well hidden 
in these coordinates. Notice that
the forms for $u$ and $v$ are effectively switched in patches I and II (with
a factor of $i=\sqrt{-1}$ removed). There is a similar switch in patches
III and IV. This is like
switching $r$ and $t$ of the standard Schwarzschild coordinates. It is not
clear what the physical meaning of such a switch would be. If $r$ is the
physical time inside a black hole, then what would be the significance of
$r=0$?

\section{The isotropic coordinates -- no black hole}
In the isotropic coordinates\cite{misner}, $r$ is replaced by $\rho$ which is defined as follows.
\begin{equation}
	r=\rho\left(1+\frac{r_{s}}{4\rho}\right)^{2}.
\end{equation}
The resulting form for the line element is
\begin{equation}
	d\tau^{2}=\left(\frac{1-\frac{r_{s}}{4\rho}}{1+\frac{r_{s}}{4\rho}}\right)^{2}dt^{2}
	-\left(1+\frac{r_{s}}{4\rho}\right)^{4}(d\rho^{2}+\rho^{2}d\Omega^{2}).
\end{equation}
\begin{figure}
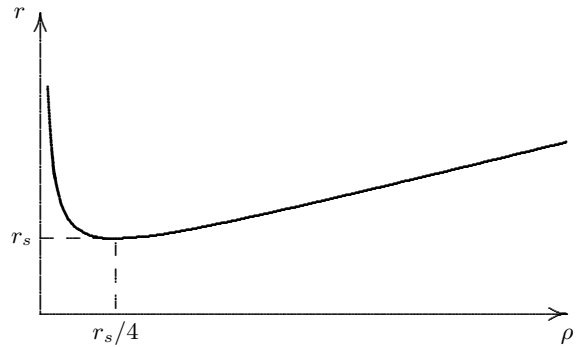

\beginpicture
\setcoordinatesystem units <1mm,1mm>
\setplotarea x from -10 to 70, y from -10 to 40
\arrow <7pt> [.2,.67] from 0 0 to 70 0
\put {$r$} [r] at -2 40
\arrow <7pt> [.2,.67] from 0 0 to 0 40
\put {$\rho$} [t] at 70 -2
\setdashes
\plot 0 10  10 10  10 0 /
\setsolid
\put {$r_{s}/4$} [t] at 10 -1
\put {$r_{s}$} [r] at -1 10
\setplotsymbol ({.})
\setquadratic
\plot
1	30.25
1.5	22.04166667
2	18
3	14.08333333
4.5	11.68055556
7	10.32142857
10	10
15	10.41666667
20	11.25
25	12.25
30	13.33333333
35	14.46428571
40	15.625
45	16.80555556
50	18
55	19.20454545
60	20.41666667
65	21.63461538
70	22.85714286
 /
\endpicture
\caption{$r$ vs.~$\rho$ the isotropic coordinate. \label{fig1}}
\end{figure}

Figure~\ref{fig1} shows a graph of $r$ vs.~$\rho$. It shows that $\rho$ does not cover
the interior of the black hole ($r<r_{s}$) at all. However, it covers the exterior twice! So,
if the range $r_{s}/4<\rho<\infty$ is defined as the physical manifold, there would be only
one coordinate patch for the complete physical manifold and the interior of the black hole
would be naturally eliminated.

However, there is one unexpected side effect of the isotropic coordinates. The smallest
sphere around the origin does not have a zero surface area! This effect will be discussed
in greater detail shortly.

\section{A class of simple coordinate systems}
The success of the isotropic coordinates in naturally eliminating the interior of the
black hole leads one to look for other simple coordinate systems that would do the same.
A class of such coordinate systems is given by the transformation from $r$ to some
$r_{n}$ defined as follows.
\begin{equation}
	r_{n}^{n}=r^{n}-r_{s}^{n},\;\;\;\mbox{for } n=1,2,3,\ldots.
\end{equation}
Each value of $n$ corresponds to a coordinate system. The range $0<r_{n}<\infty$ is completely
outside the black hole. This may seem like a rather ad hoc way of eliminating the
interior of the black hole. However, it is interesting to note that in Schwarzschild's
original work\cite{schwarz, schwarzt} the physical radial coordinate used was not $r$ but 
$r_{3}$ as given by the $n=3$ case of the above class of coordinate systems.

Such ad hoc coordinate systems are given legitimacy by the unphysical nature of the
interior of the black hole. No probe can be sent by an outside observer to a black hole 
to examine its interior and return with the information. A quantum theoretical analysis
has also produced similar results\cite{vachas}.

Hence, a closer look at this class of coordinate systems is warranted. The $n=1$ case is
quite trivial and the $n=3$ case is already known from Schwarzschild's work. So, let us look
at the $n=2$ case. For $n=2$ we get
\begin{equation}
	r_{2}^{2}=r^{2}-r_{s}^{2}.
\end{equation}

\begin{figure}
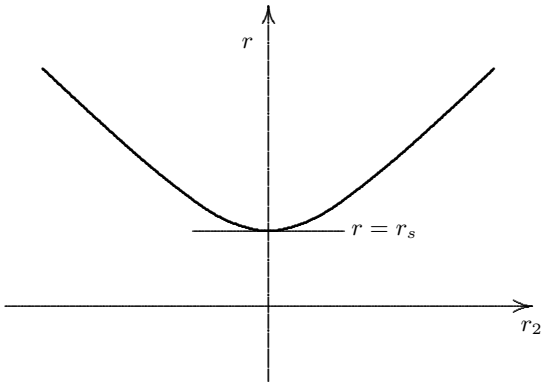

\beginpicture
\setcoordinatesystem units <1mm,1mm>
\setplotarea x from -35 to 35, y from -10 to 45
\arrow <7pt> [.2,.67] from -35 0 to 35 0
\put {$r$} [r] at -2 35
\arrow <7pt> [.2,.67] from 0 -10 to 0 40
\put {$r_{2}$} [t] at 35 -2
\plot -10 10  10 10 /
\put {$r=r_{s}$} [l] at 11 10
\setplotsymbol ({.})
\setquadratic
\plot
-30	31.6227766
-25	26.92582404
-20	22.36067977
-15	18.02775638
-10	14.14213562
-5	11.18033989
0	10
5	11.18033989
10	14.14213562
15	18.02775638
20	22.36067977
25	26.92582404
30	31.6227766
 /
\endpicture
\caption{$r$ vs.~$r_{2}$: a hyperbolic coordinate. \label{fig2}}
\end{figure}

Figure~\ref{fig2} shows the $r$ vs.~$r_{2}$ graph. Qualitatively, this is similar to the
case of isotropic coordinates. The minimum value of $r$ is the Schwarzschild radius $r_{s}$.
But unlike $\rho$ of the isotropic coordinates, $r_{2}=0$ at the minima. Physically, it is
possible to define $r_{2}$ in the range $-\infty<r_{2}<+\infty$. As a result, it covers
the physical space twice.

The form for the line element is
\begin{eqnarray}
d\tau^{2} & = & \left(1-\frac{r_{s}}{(r_{2}^{2}+r_{s}^{2})^{1/2}}\right)dt^{2} \nonumber \\
          &   & -(r_{2}^{2}+r_{s}^{2}-r_{s}(r_{2}^{2}+r_{s}^{2})^{1/2})^{-1}r_{2}^{2}dr_{2}^{2}
          \nonumber \\
          &   & -(r_{2}^{2}+r_{s}^{2})d\Omega^{2}.
\end{eqnarray}

Once again the smallest sphere around the origin does not
have zero surface area. So, next we shall consider the physical acceptability of this
phenomenon.

\section{The surface area of spheres around the origin}
Historically, it is not clear why Schwarzschild's original radial coordinate was abandoned
in favor of the currently accepted $r$. It is very likely because the Schwarzschild radial
coordinate (called $r_{3}$ here\footnote{What is called $r_{3}$ here was called $r$ by
Schwarzschild. Here, we call it $r_{3}$ to distinguish it from the currently accepted $r$.}) 
has a rather unusual feature -- the surface area
of the smallest spherical surface around the origin ($r_{3}=0$) is not zero! However, one
needs to note that the system under consideration here is also unusual. Gravity is expected
to be very strong near the horizon and hence space is expected to be significantly distorted
due to curvature. So, the usual relation of surface area $S$ and radius $r$ of a sphere
should not be expected. However, in standard coordinates this is exactly what is assumed:
\begin{equation}
	S=4\pi r^{2}. \label{eqsphsur}
\end{equation}
This assumption is built into the choice of the angular part of the metric -- $r^{2}d\Omega^{2}$.
Note that in a standard textbook presentation of the derivation of the Schwarzschild metric the
following ansatz is used for the metric.
\begin{equation}
	d\tau^{2}=A(r)dt^{2}-B(r)dr^{2}-r^{2}d\Omega^{2},
\end{equation}
The undetermined functions $A(r)$ and $B(r)$ are determined from Einstein's equations.
Schwarzschild used an ansatz with three undetermined functions as follows.
\begin{equation}
	d\tau^{2}=A(r)dt^{2}-B(r)dr^{2}-C(r)r^{2}d\Omega^{2},
\end{equation}
This does not agree with equation~\ref{eqsphsur} unless $C(r)=1$. The standard 
argument for choosing $C(r)=1$ is the freedom of coordinate choice for $r$.
Although such a choice is legitimate, it results in a loss of the physical meaning of 
the origin of $r$. There is no reason to believe that the physical manifold starts at
$r=0$ and not at $r=r_{s}$. Schwarzschild assumed the physical manifold starts at 
$r_{3}=0$ which is the same as $r=r_{s}$. In fact, all coordinate systems defined in the
last two sections have this property.

To appreciate the significance of not assuming equation~\ref{eqsphsur} for curved spaces, 
one can consider some simple 2-dimensional
curved surfaces. The circumference $C$ of a circle drawn on such a surface can be related
to the radius $r$ in very different ways compared to flat space. For example, a circle
drawn on a 2-dimensional conical surface with the apex as center has a circumference of
\begin{equation}
	C=(2\pi\sin\theta)r,
\end{equation}
where $\theta$ is the half angle of the cone and $r$ is the radius of the circle.
Another example is that of a circle drawn on a 2-dimensional spherical surface. Here
the circumference is related to radius as follows.
\begin{equation}
	C=2\pi R\sin(r/R),
\end{equation}
where $R$ is the radius of the spherical surface. This relationship of $C$ and $r$
is not even linear. Hence, it should not be surprising to find the surface area of
a sphere in 4-dimensional curved space-time to be related to the radial coordinate
in ways different from the flat space case given in equation~\ref{eqsphsur}.

\section{Conclusion}
It is customary to assume that the standard radial coordinate $r$ used for the
Schwarzschild metric has the physical range $0<r<\infty$. Here it is shown that
there is no fundamental reason for assuming this range. Indeed, there exist
``natural'' coordinate systems that exclude the region inside the black hole.
Such coordinates consider the physical range to be $r_{s}<r<\infty$. Besides, there 
are strong physical reasons for excluding the interior of the black hole. One
reason being the switching of the metric signature within the black hole. Another
is the physical inaccessibility of interior black hole information to the outside
observer. This makes it impossible for an outside observer to determine for sure
if an object is indeed a black hole.

\end{document}